# Performant Dynamically Typed E-Graphs in Pure Julia

Unleashing Equality Saturation in a Modern Language for Scientific Computing

Alessandro Cheli
PlantingSpace
alessandro@cheli.dev
Github: `@0x0f0f0f`

Niklas Heim
Czech Technical University
heimnikl@fel.cvut.cz
Github: `@nmheim`

**Abstract**

We introduce the third major version of Metatheory.jl, a Julia library for general-purpose metaprogramming and symbolic computation. Metatheory.jl provides a flexible and performant implementation of e-graphs and Equality Saturation (EqSat) that addresses the two-language problem in high-level compiler optimizations, symbolics and metaprogramming. We present results from our ongoing optimization efforts, comparing the state-of-the-art egg Rust library's performance against our system and show that performant EqSat implementations are possible without sacrificing the comfort of a direct 1-1 integration with a dynamic, high-level and an interactive host programming language.

## 1. Introduction

**Why Julia: The Two-Language Problem.** Domain-agnostic Equality Saturation (EqSat) and e-graph term rewriting techniques have seen substantial growth and recognition in the past years, originating from the influential egg [1] library. While implementations in system programming languages such as Rust showcase excellent performance, most scientists are accustomed to high-level programming languages such as Python, Mathematica and MATLAB. The need of integrating performant programs and easy to use interfaces arises: scientists prototype algorithms in high-level languages, but are subsequently compelled to rewrite their program in a lower-level language. This is time-consuming and error-prone, and often necessitates the evolution from individual expertise to specialized team collaboration. This is known in literature as the *two-language problem*. Julia [2], [3] offers an interactive, user-friendly, performant and compiled language, thus solving this problem in many practical applications [2]–[7] and eliciting positive feedback from users [8]. This has fostered a collaborative and interdisciplinary community around the language and its open-source ecosystem, driving the development of numerous state-of-the-art scientific computing and symbolic-numeric applications. Julia encourages writing functional and composable code, where shared interfaces are re-used across the open-source package ecosystem. This is possible thanks to *multiple dispatch*, a feature where functions can possess multiple executable methods, each associated with a different combination of input argument types.

**Julia Metaprogramming.** Julia provides *excellent metaprogramming features*: the language retains an interactive shell, *true homoiconicity*, Scheme-like *hygienic macros*, and interactive utilities to inspect intermediate representations of code, all while being a compiled language. It is thus remarkably easy to extend the language with *custom eDSLs* (Embedded Domain-Specific Languages) that can be either interpreted at compilation time or interpolated back in regular Julia programs.

**Introducing Metatheory.jl**. The features listed above make Julia an excellent candidate language for *equality saturation workflows*, as performant numerical programs can be optimized, synthesized or rewritten within themselves, at both compilation and execution times. Practically solving the two-language problem in the context of EqSat workflows.

Our open-source package Metatheory.jl[1] [9], [10] is a Julia library for general purpose algebraic metaprogramming and symbolic computation, leveraging innovative features of Julia to provide a flexible, user friendly and performant implementation of *e-graphs* and EqSat, fully integrated with the language's dynamical type system through a *rewrite rule eDSL and pattern matcher*, that is *backwards-compatible* and *easily integrated* with existing symbolic computation packages such as Symbolics.jl. [11]. At the time of writing Metatheory.jl is implemented in approximately 13000 lines of code, inclusive of a comprehensive test suite, Continuous Integration (CI) benchmarking, example rewrite systems and narrative tutorials in literate programming style.

## 2. Features of Metatheory.jl

**A shared interface for tree-like objects.** In Julia metaprogramming, programs are stored in the tree-shaped `Expr` data structure, akin to Lisp S-expressions. In contrast, complex symbolic algebra frameworks may necessitate the definition of *custom expression data types*. One of the main challenges in the development of Metatheory.jl, was designing a shared expression tree interface that supported both Symbolics.jl [11] terms, Julia code and custom data types. We introduce TermInterface.jl [12], an open-source package providing such interface. To use Metatheory.jl rewriting features on an custom expression data type `T`, a user should implement methods for the following functions:

- **`isexpr(x::T)`** returns `true` $\iff$ `x` is an expression tree (an S-expression). If true, `head` and `children` methods must be defined for `x`.
- **`iscall(x::T)`** returns `true` $\iff$ `x` is a function call expression. If true, `operation` and `arguments` must also be defined for x::T. Note that `iscall(x)` $\implies$ `isexpr(x)`, but not the other way around.
- **`head(x) and children(x)`** behave like Lisp's `car` and `cdr`, returning head and tail of the expression.
- **`operation(x) and arguments(x)`** return the function and arguments a of function call expression only when `iscall(x)` is true. These functions were introduced because some expression types, such as Julia `Expr`, do not store the function being called as the head of the term.
- **`maketerm(T, head, children, type=nothing, metadata=nothing)`** constructs an expression where `T` is a constructor type, usually the type of the input expression which is being manipulated. For example, when a subexpression is substituted, the outer expression is re-constructed with the sub-expression. `T` will be the type of the outer expression. The optional argument `type` can be used to attach type information (e.g. when representing typed lambda calculi), and `metadata` to attach additional information.

**Unified Rule eDSL**. Our package offers an *unified rewrite rule definition eDSL* that is backwards-compatible with the computer-algebra-oriented classical rewriting package SymbolicUtils.jl [11]. We strive for a pattern eDSL that is user friendly and as visually close as possible to mathematics in literature. In Listing 1 we show the three main kind of rewrite rules that are available in our library, and how the `@rule` macro expands to construct a rule that captures the scope around its callsite. Our pattern matching and e-matching engines compile patterns directly to Julia closures, similarly to Continuation Passing Style (CPS) compilation. There is no need of a virtual machine for backtracking [1], [13] as the Julia function call stack is used directly. Although there is room for further performance optimizations [14], our approach ensures efficient integration with multiple dispatch and the Julia type system, as can be seen in Listing 2. Our rule language supports three different rule types: directed, bidirectional (equational) and dynamic rule types, where the right-hand-side is efficiently evaluated as Julia code. We support conditional rules and predicates, that thanks to multiple-dispatch can efficiently match distinguishing expression trees, constant literals and e-classes. As rules can be dynamically constructed at runtime, our eDSL also offers a convenient framework for rewrite rule synthesis and inference via EqSat, techniques that have already explored with egg [15].

---
[1]Source code and documentation is available at https://github.com/JuliaSymbolics/Metatheory.jl

```julia
1  # Equational Rewrite Rule
2  julia> r =  @rule ~a * (~b + ~c) ==
3                      ((~a * ~b) + (~a * ~c));
4  # Dynamic Rewrite Rule. RHS is evaluated.
5  # Useful for constant propagation without analyses.
6  julia> dr = @rule ~a::Number ⊗ ~b::Number => a*b;
7  # Directed rewrites work in classical rewriting.
8  julia> cr = @rule ~a + 0 --> ~a;
9
```

User-level interface for rule definition.

```julia
1  julia> dr(:(2.0 ⊗ 4.3))
2  8.6
3  julia> cr(:((x * y) + 0))
4  :(x * y)
5  julia> r(:(x * (y + z)))
6  ERROR: MethodError: objects of type EqualityRule
7    are not callable
```

Rules are callable objects for classical rewriting.

```julia
1  julia> @macroexpand @rule ~a::Number ⊗
2                            ~b::Number => a*b
3  # The result of macro expansion is quoted code
4  quote
5      DynamicRule(
6        # Expansion of LHS with automatic quoting
7        # of undefined function names.
8        PatExpr(true,
9          isdefined(Main, :*) ? * : (:*),
10           PatVar(:a, -1, Number, :Number),
11           PatVar(:b, -1, Number, :Number)
12       ),
13       # RHS of `=>` rules is expanded to
14       # an anonymous function definition.
15       ((_lhs_expr, _egraph, a, b)->begin
16         a * b
17       end),
18       # The original expression of the RHS
19       # is preserved for printing
20       $(QuoteNode(:(a * b))))
21  end
```

After Macro Expansion

Listing 1: Example of Metatheory.jl rewrite rules definitions via the `@rule` macro. Macros are interpolated back into the source program, automatically detecting what objects are defined at runtime.

```julia
1  abstract type Vehicle end
2  abstract type GroundVehicle <: Vehicle end; abstract type AirVehicle <: Vehicle end
3  struct Airplane <: AirVehicle end; struct Car <: GroundVehicle end
4  airpl, car = Airplane(), Car()
5  t = [@rule(~a::AirVehicle => "flies"); @rule(~a::GroundVehicle => "drives")]
6  Chain(t)(car) # returns "drives"
```

Listing 2: Example showing how our eDSL integrates with the Julia type system.

**Classical Rewriting, Functional Combinators and Pattern Matching**. In both Symbolics.jl [11] and Metatheory.jl [9], rewrite rules are *callable objects*, where the only argument is the expression tree to pattern match and eventually substitute. Intuitively, bidirectional rules (==) are not supported in a classical rewriting context, but dynamic (=>) and rewrite rules (-->) behave as expected. We offer a collection of *functional rewriter combinators*, such as `Chain`, `Fixpoint`, `Prewalk` and `Postwalk`, allowing users to compose rewrite systems for classical rewriting workflows in a functional programming style. This creates the opportunity to compose classical rewriting with equality saturation in separate steps, using the same rewrite system definitions.

**Integration with Symbolics.jl and the SciML ecosystem** *SciML* (Scientific Machine Learning) is a Julia-centered organization dedicated to advancing research in both numerical simulation and scientific machine learning. SciML's packages have been widely employed in a range of scientific applications: Climate Modeling, Pharmaceutical Simulation, Space Research and many more [16]. State-of-the-art systems such as NeuralPDE.jl [17] (Physics-informed neural networks) are not rare encounters in the SciML software ecosystem.

One of the key packages in the SciML ecosystem is ModelingToolkit.jl [18], extending Julia with a modeling language for high-performance symbolic-numeric computation, integrating the Symbolics.jl [11] CAS (Computer Algebra System) with causal and acausal equation-based modeling frame-works and state-of-the-art differential equation solvers [6]. We previously showed how Metatheory.jl can be used to apply EqSat on symbolic expressions generated by ModelingToolkit.jl and Symbolics.jl before numerical evaluation, halving the time required by a chemical reaction network simulation (highly stiff ordinary differential equations) and obtaining a 8x speedup when numerically evaluating a symbolic mass matrix of a rigid body dynamics system [10]. With the newly released TermInterface.jl version [12], deeper integration between Symbolics.jl and Metatheory.jl is possible, opening up new opportunities in using EqSat to accelerate state-of-the-art symbolic-numeric workflows.

**Automatic GraphViz Package Extension for e-graph visualization.** Julia 1.9's package manager introduces *weak dependencies* through *"package extensions"*. If GraphViz.jl is installed on a user's system when loading our package, Julia will automatically load methods to visualize e-graphs as vector image via GraphViz [19], enhancing compatibility with development environments that support image display, such as Visual Studio Code, Jupyter and Pluto notebooks.

## 3. Practical Example and Performance Evaluation

**Example Equality Saturation Workflow.** While egg's syntax requires parsing rewrite rules from S-Expression strings into objects, we strive for an user interface that can fully interface with the existing Julia parser and macro system, remaining simple, intuitive and allowing researchers to write code that resembles mathematics. In Listing 3 we show a comparison of two identical EqSat workflows, formulated in both Rust (egg) and Julia (Metatheory.jl). In Listing 4 we show how Metatheory.jl can be used for custom, high-level compiler optimizations based on domain-specific equational theories. The same rewrite system can be integrated seamlessly with the Julia macro system, rewriting code via EqSat before the actual function is compiled.

**Performance evaluation.** We compared the performance of EqSat in Metatheory.jl and egg on some minimal e-graph rewrite systems that require constant propagation and equational reasoning.[2] In Table 1 we report the median execution time of such benchmarks and comparisons between libraries. The benchmarking time includes e-graph creation, one or multiple iterated executions of EqSat and extraction. Metatheory.jl's implementation of EqSat is closely modeled after egg and the benchmarks include equivalent systems of rewrites, EqSat parameters and halting conditions. For the sake of simplicity, we are always using an e-graph extraction cost function that chooses the smallest terms in e-classes (`astsize`). Readers can refer to the original egg paper for an accurate description of the e-graph extraction process [1]. The first column reports the name of the benchmark, while the second and third columns report the median execution time in egg. The *egg-sym* column refers to benchmarks implemented using `SymbolLang`, a generic tree structure, while *egg-cust* benchmarks were implemented with `define_language!`, generating optimized expression data structures. Columns *MT2* and *MT3* report the execution times in Metatheory.jl version 2.0 and the currently in-development version 3.0. Metatheory.jl benchmarks were implemented using the internal Julia expression type `Expr`. The remaining columns report a ratio, comparing how Metatheory.jl 3.0 performs against its previous release and egg. As Metatheory.jl can rewrite on any Julia data type that implements TermInterface.jl [12], there is a tradeoff between dynamism and speed, as substantial overhead is indeed introduced by repeated construction and garbage collection of such a generic data structure. We intensively optimized our package during development of its third major release: we obtained speedups up to $\times 226$ on simple propositional theorem proving tasks, and averaging over the benchmarks, we were able to obtain an impressive $\times 41$ speedup in the time required by equality saturation. One of the reasons why were able to achieve such performance gains was switching from custom e-node data structures to compact vectors of unsigned integers, as the Julia compiler provides excellent optimizations such as automatic vectorization of loops.

---

[2]Code for benchmarks is available at https://github.com/nmheim/egg-benchmark/. The results were obtained on Standard GitHub-hosted runners for Public repositories, 16GB RAM, using Julia 1.11, Metatheory.jl 3.0, rustc 1.77.1 and egg 0.9.5. Similar results were obtained on x86-64 Linux machines on identical software versions.

| Benchmark Name | egg-sym | egg-cust | MT3 | MT2 | egg-sym/MT3 | egg-cust/MT3 | MT2/MT3 |
|---|---|---|---|---|---|---|---|
| egraph_addexpr | 1.45 ms | | 4.49 ms | 15.3 ms | 0.323 | | 3.4 |
| basic_maths_simpl2 | 14.5 ms | 5.12 ms | 17.5 ms | 884 ms | 0.827 | 0.292 | 50.4 |
| prop_logic_freges_theorem | 2.7 ms | 1.56 ms | 2.24 ms | 35.7 ms | 1.21 | 0.698 | 15.9 |
| calc_logic_demorgan | 65.3 µs | 34.2 µs | 82.8 µs | 514 µs | 0.788 | 0.412 | 6.21 |
| calc_logic_freges_theorem | 23.4 ms | 12.1 ms | 35.9 ms | 3.24e+03 ms | 0.652 | 0.336 | 90.3 |
| basic_maths_simpl1 | 6.91 ms | 2.8 ms | 5.21 ms | 50.8 ms | 1.33 | 0.538 | 9.76 |
| egraph_constructor | 0.0851 µs | | 0.0921 µs | 0.105 µs | 0.924 | | 1.14 |
| prop_logic_prove1 | 37.1 ms | 14.1 ms | 39.8 ms | 8.97e+03 ms | 0.933 | 0.354 | 226 |
| prop_logic_demorgan | 86.1 µs | 45.5 µs | 101 µs | 758 µs | 0.856 | 0.452 | 7.53 |
| Package time to load | | | 503 ms | 541 ms | | | 1.08 |
| Average Ratio | | | | | 0.871 | 0.44 | 41.17 |

Table 1: Performance comparison between egg and Metatheory.jl

```
1  use egg::{rewrite as rw, *};
2  fn main() {
3      env_logger::init();
4      use egg::*;
5      define_language! {
6          enum SimpleLanguage {
7              Num(i32),
8              "+" = Add([Id; 2]),
9              "*" = Mul([Id; 2]),
10             Symbol(Symbol),
11         }
12     }
13     fn make_rules() -> Vec<Rewrite<SimpleLanguage, ()>> {
14         vec![
15             rewrite!("commute-add"; "(+ ?a ?b)" => "(+ ?b ?a)"),
16             rewrite!("commute-mul"; "(* ?a ?b)" => "(* ?b ?a)"),
17             rewrite!("add-0"; "(+ ?a 0)" => "?a"),
18             rewrite!("mul-0"; "(* ?a 0)" => "0"),
19             rewrite!("mul-1"; "(* ?a 1)" => "?a"),
20         ]
21     }
22     fn simplify(s: &str) -> String {
23         let expr: RecExpr<SimpleLanguage> = s.parse().unwrap();
24         let runner = Runner::default().with_expr(&expr).run(&make_rules());
25         let root = runner.roots[0];
26         let extractor = Extractor::new(&runner.egraph, AstSize);
27         let (best_cost, best) = extractor.find_best(root);
28         println!("Simplified {} to {} with cost {}", expr, best, best_cost);
29         best.to_string()
30     }
31
32     let apply_time: std::time::Instant = instant::Instant::now();
33     assert_eq!(simplify("(+ (+ (+ 0 (* (* 1 x) 0)) (* y 0)) y)"), "y");
34     let apply_time = apply_time.elapsed().as_secs_f64();
35     println!("simplification time {}", apply_time);
36 }
```

Rust, using egg.

```
1  using Metatheory, BenchmarkTools
2  t = @theory a b begin
3    # Equational Rules
4    a + b == b + a
5    a * b == b * a
6    # Directed rewrites
7    a + 0 --> a
8    a * 1 --> a
9    # Dynamic Rules: RHS is evaluated as Julia
10   a * 0 => 0
11 end
12 p = SaturationParams(; timeout = 8, timer = false)
13 function simpl(ex)
14   g = EGraph(ex)
15   saturate!(g, t)
16   extract!(g, astsize)
17 end
18 ex = :(0 + (1 * x) * 0 + (y * 0) + y)
19 @assert simpl(ex) == :y
20 @btime simpl(ex)
```

Julia, using Metatheory.jl.

Listing 3: Comparison of an EqSat workflow in Rust using egg, rewriting a custom expression type that has to be defined in Rust, and in Julia, using Metatheory.jl, rewriting the homoiconic `Expr` type.

```
1  macro simplify(e::Expr)
2      esc(simpl(e))
3  end
4  unoptimized(x,y) = (log(x) * exp(x) / x^5) * 0 + y
5  optimized(x,y) = @simplify (log(x) * exp(x) / x^5) * 0 + y
```

Listing 4: Integrating Listing 3 in a macro, optimizing the program with EqSat before compilation.

# 4. In-Development and Future Work

**Proof-production.** Given a system of bidirectional and directed rewrite rules, proving the equality of two expressions $a$ and $b$ via EqSat requires an expensive, non-deterministic e-graph search [13], [1]. A corresponding proof certificate can be intuitively represented by a chain of directed-only rewrites that can be applied sequentially to rewrite expression $a$ to $b$ in linear time (or vice-versa), without the need of an expensive e-graph search. This functionality is crucial in theorem proving environments that rely on congruence closure algorithms, such as Satifiability Modulo Theory (SMT) solvers. In egg it is possible to generate proof certificates via a feature named *explanations*: a state-of-the-art algorithm generating optimal proofs from an e-graph in $O(n^5)$ time [20]. Implementing a similar feature in Metatheory.jl will lay the foundations for a pure Julia interactive theorem proving environment, helping scientists to reason about complex, domain-specific theories that are equational by nature and are commonly used to develop symbolic-numeric simulations [16].

**Enhanced Scheduling Algorithms and Parameter Inference.** E-matching [13] every rewrite rule against the entire e-graph is costly. Both egg and our system provide *every-rule-every-time* and *exponential backoff* heuristics called *schedulers* [1]. The latter identifies rewrites that produce an exponential number of matches and temporarily bans them. While this is effective for many EqSat workflows, we aim to develop *goal-informed* heuristics, effectively reducing the algorithm's search space and execution time by selectively filtering rewrites in each EqSat iteration, based on the users' intent, such as minimizing a cost function or proving equalities. All of the ingredients are present to formulate scheduling as a *Reinforcement Learning* problem. Other attempts to integrate EqSat with such techniques were attempted in [21] and [22]. To further improve EqSat performance, automatic hyperparameter inference techniques can be developed.

**Optimizing Floating Point Expressions.** We are currently developing OptiFloat.jl, a pure Julia floating point expression optimizer, heavily inspired by Herbie [23]. The project's goal is to improve accuracy as well as performance of floating point expressions. Much of Julia's SciML ecosystem heavily relies on purely numeric, symbolic expressions, which would immediately benefit from OptiFloat.jl without having to rely on Julia wrappers for external systems implementend in Racket or Rust. The strong composable nature of our package OptiFloat.jl will enable optimization of symbolic expressions, native Julia code and any custom expression type implementing TermInterface.jl [12]. With all the existing infrastructure provided by Metatheory.jl, the core of this floating point expression optimization amounts to appropriate cost function definitions and extraction algorithms.

**Typed Expressions.** The concept of a symbolic type is already supported by Symbolics.jl to represent the return type of an expression. It is possible to extend our rule eDSL and pattern matcher to efficiently support aribtrary *typed expressions*, as in typed calculi. There are ongoing discussions about extending the eDSL syntax by introducing a triple column operator, where `g(~a, ~b)` is a pattern that will match against the type, `f(~x, ~y):::(g(~a, ~b))` would pattern match against a typed expression. The double column operator is already used for types and predicates in patterns such as `~x::Number`.

**Further Extensions.** Thanks to *multiple dispatch*, extensions to the core data structures and algorithms can be prototyped in a straightforward manner. For example, implementing *colored e-graphs* [24] in Metatheory.jl would require the extension of *only* the core EqSat functions where the new data type requires changes. The Julia open-source ecosystem contains packages that provide Prolog-style reasoning engines in eDSLs [25]. Metatheory.jl may serve as the foundational infrastructure for a pure Julia implementation of `egglog` [14], a fixpoint reasoning system that unifies Datalog and equality saturation.

# Bibliography

[1] M. Willsey, C. Nandi, Y. R. Wang, O. Flatt, Z. Tatlock, and P. Panchekha, "Egg: Fast and extensible equality saturation," *Proceedings of the ACM on Programming Languages*, vol. 5, no. POPL, pp. 1–29, 2021.

[2] J. Bezanson, S. Karpinski, V. B. Shah, and A. Edelman, "Julia: A fast dynamic language for technical computing," *arXiv preprint arXiv:1209.5145*, 2012.

[3] J. Bezanson, A. Edelman, S. Karpinski, and V. B. Shah, "Julia: A fresh approach to numerical computing," *SIAM review*, vol. 59, no. 1, pp. 65–98, 2017.

[4] T. Koolen and R. Deits, "Julia for robotics: Simulation and real-time control in a high-level programming language," in *2019 International Conference on Robotics and Automation (ICRA)*, 2019, pp. 604–611.

[5] E. Roesch *et al.*, "Julia for biologists," *Nature methods*, vol. 20, no. 5, pp. 655–664, 2023.

[6] C. Rackauckas and Q. Nie, "Differentialequations.jl–a performant and feature-rich ecosystem for solving differential equations in julia," *Journal of Open Research Software*, vol. 5, no. 1, p. 15–16, 2017.

[7] M. Lindner *et al.*, "NetworkDynamics. jl—Composing and simulating complex networks in Julia," *Chaos: An Interdisciplinary Journal of Nonlinear Science*, vol. 31, no. 6, 2021.

[8] A. Claster, "2023 Julia User & Developer Survey." [Online]. Available: https://julialang.org/assets/2023-julia-user-developer-survey.pdf

[9] A. Cheli, "Metatheory.jl: Fast and Elegant Algebraic Computation in Julia with Extensible Equality Saturation," *Journal of Open Source Software*, vol. 6, no. 59, p. 3078–3079, 2021, doi: 10.21105/joss.03078.

[10] A. Cheli, "Automated Code Optimization with E-Graphs," *arXiv preprint arXiv:2112.14714*, 2021.

[11] S. Gowda *et al.*, "High-Performance Symbolic-Numerics via Multiple Dispatch," *ACM Commun. Comput. Algebra*, vol. 55, no. 3, pp. 92–96, Jan. 2022, doi: 10.1145/3511528.3511535.

[12] JuliaSymbolics, "TermInterface.jl." [Online]. Available: https://github.com/JuliaSymbolics/TermInterface.jl

[13] L. De Moura and N. Bjørner, "Efficient E-matching for SMT solvers," in *Automated Deduction–CADE-21: 21st International Conference on Automated Deduction Bremen, Germany, July 17-20, 2007 Proceedings 21*, 2007, pp. 183–198.

[14] Y. Zhang *et al.*, "Better together: Unifying datalog and equality saturation," *Proceedings of the ACM on Programming Languages*, vol. 7, no. PLDI, pp. 468–492, 2023.

[15] C. Nandi *et al.*, "Rewrite rule inference using equality saturation," *Proceedings of the ACM on Programming Languages*, vol. 5, no. OOPSLA, pp. 1–28, 2021.

[16] "SciML Scientific Machine Learning Showcase." [Online]. Available: https://sciml.ai/showcase/

[17] S. Cuomo, V. S. Di Cola, F. Giampaolo, G. Rozza, M. Raissi, and F. Piccialli, "Scientific machine learning through physics–informed neural networks: Where we are and what's next," *Journal of Scientific Computing*, vol. 92, no. 3, p. 88–89, 2022.

[18] Y. Ma, S. Gowda, R. Anantharaman, C. Laughman, V. Shah, and C. Rackauckas, "ModelingToolkit: A Composable Graph Transformation System For Equation-Based Modeling." 2021.

[19] J. Ellson, E. Gansner, L. Koutsofios, S. C. North, and G. Woodhull, "Graphviz—open source graph drawing tools," in *Graph Drawing: 9th International Symposium, GD 2001 Vienna, Austria, September 23–26, 2001 Revised Papers 9*, 2002, pp. 483–484.

[20] O. Flatt, S. Coward, M. Willsey, Z. Tatlock, and P. Panchekha, "Small Proofs from Congruence Closure," in *2022 Formal Methods in Computer-Aided Design (FMCAD)*, 2022, pp. 75–83.

[21] G.-O. Barbulescu, "An Inquiry into Database Query Optimisation with Equality Saturation and Reinforcement Learning," 2023.

[22] Z. Singh, "Deep Reinforcement Learning for Equality Saturation," 2022.

[23] P. Panchekha, A. Sanchez-Stern, J. R. Wilcox, and Z. Tatlock, "Automatically improving accuracy for floating point expressions," *Acm Sigplan Notices*, vol. 50, no. 6, pp. 1–11, 2015.

[24] E. Singher and S. Itzhaky, "Colored E-Graph: Equality Reasoning with Conditions," *arXiv preprint arXiv:2305.19203*, 2023.

[25] ztangent, "Julog.jl." [Online]. Available: https://github.com/ztangent/Julog.jl